\UseRawInputEncoding 
\documentclass[journal]{IEEEtran}

\usepackage{xcolor}

\usepackage{siunitx}

\ifCLASSINFOpdf
  \usepackage[pdftex]{graphicx}
\else
\fi
\begin{document}
%
\title{Antenna Radiation Efficiency Estimation From Backscattering Measurement Performed Within Reverberation Chambers}
%
%
%

\author{Wafa~Krouka,
        Fran\c{c}ois~Sarrazin,~\IEEEmembership{Member,~IEEE,}
        Julien~de~Rosny,~\IEEEmembership{Senior~Member,~IEEE,}
        Adnane~Labdouni,
        and~Elodie~Richalot,~\IEEEmembership{Member,~IEEE}
\thanks{W. Krouka, F. Sarrazin, A. Labdouni and E. Richalot are with the ESYCOM lab, Univ Gustave Eiffel, CNRS, F-77454 Marne-la-Vallée, France (e-mail: francois.sarrazin@univ-eiffel.fr).}
\thanks{J. de Rosny is with ESPCI Paris, PSL University, CNRS, Institut Langevin, Paris, France (julien.derosny@espci.fr).}}

%
%

\markboth{This work has been submitted to the IEEE for possible publication. Copyright may be transferred without notice.}%
{Shell \MakeLowercase{\textit{et al.}}: Bare Demo of IEEEtran.cls for IEEE Journals}
%

\maketitle

\begin{abstract}
This paper presents a contactless measurement method for antenna radiation efficiency estimation within reverberation chambers (RCs). This method does not require to connect any analyzer to the antenna under test (AUT). The AUT radiation efficiency is obtained from the differential measurement of the RC composite $Q$-factor estimated for two AUT load conditions, namely an open circuit and a \SI[detect-weight]{50}{\ohm} load.  Remotely-controlled RF switches are used in order to keep the setup identical for both measurements. The method is experimentally validated in two RCs of different volumes (1~m$^3$ and 19~m$^3$) with a wideband patch antenna in the  1.8~GHz to 2.8~GHz  frequency range. The estimated efficiency is found to be in very good agreement with the one obtained through a conventional invasive RC measurement technique. This new method is suitable for miniature and buried antenna characterization without access to its port.
\end{abstract}

\begin{IEEEkeywords}
Antenna radiation efficiency, backscattering measurement, contactless measurement, quality factor, reverberation chamber.
\end{IEEEkeywords}

\IEEEpeerreviewmaketitle

\section{Introduction}

\IEEEPARstart{A}{ntenna} radiation efficiency is defined as the ratio between the radiated power and the incident power, i.e., the one that goes through the antenna port \cite{Balanis2005}. Its measurement has been initially performed using a transmission-type setup within an anechoic chamber (AC) by applying the gain-integration technique. Since first studies in 2001 \cite{Hallbjorner2001}, reverberation chambers (RCs) became very attractive to measure the antenna radiation efficiency \cite{Piette2004}--\cite{Krouka2018}. Indeed, thanks to the statistically isotropic and homogeneous field within the RC working volume, such measurement neither exhibits alignment constraints nor require the antenna under test (AUT) rotation. Moreover, the use of a reference antenna can be avoided by comparing the RC $Q$-factor estimated in the time and frequency domains \cite{Holloway2012,Krouka2020}. However, all current methods require an invasive setup, i.e., a setup where the AUT needs to be connected to the analyzer implying connectors and long cables.

The presence of cables in the measurement setup leads to unwanted interactions between the AUT and the cables \cite{Demarinis1988, Saario1997}. Two different phenomena occur: 1) the feed cables are placed in the antenna near-field zone; therefore both resistive and reactive parts of the input impedance are modified, as well as the antenna radiation characteristics, and 2) The imperfect balance between the radiating element and the ground plane causes current leaking on the outer shielding of the feed cables, so that the cables themselves act as radiators \cite{Icheln1999}--\cite{Staub1998}. These two effects become critical when dealing with electrically small antennas and lead to inaccurate results \cite{Huitema2014, Leong2007}.

Different ways have been investigated to overcome this issue in an AC: limiting the cable effects by adding ferrites, quarter-wavelength sleeves \cite{Icheln2000} or balun \cite{Fukasawa2019}; compensating for disturbancies introduced by the measurement setup thanks to prior electromagnetic simulations \cite{Huitema2014}  or post-processing \cite{Araque2011} ; replacing the coaxial cables by optical links \cite{Lao2005, Hachemi2010}; performing antenna backscattering measurement. The latter is of particular interest as it enables a contactless measurement,  i.e., without the need to connect the AUT to an analyzer. It is based on several measurements with either a varying load \cite{Appel1979} or a few discrete ones \cite{Wiesbeck1998}. First introduced in an AC, such techniques have been recently applied within an RC \cite{Reis2021}--\cite{Reis2022}  but only for antenna radiation pattern estimation. 

The insertion of lossy objects within an RC has been studied in terms of absorption \cite{Carlberg2004} and diffusion \cite{Lerosey2007} whereas the specific case of antennas has been first dealt with as a dissipative load impedance \cite{Hill1998} before that a refined model was proposed recently \cite{Cozza2018} in order to take into account the reflections due to the load impedance mismatch.
In this paper, we take benefit from these studies to introduce an original contactless measurement technique that allows retrieving the antenna radiation efficiency.

This paper is organized as follows: Section~\ref{sec:theory} introduces the theoretical background for the contactless measurement method. Section~\ref{sec:esycom} describes the experimental setup within an RC as well as the radiation efficiency results obtained using the contactless method and compared to conventional measurements. The very good agreement between the two methods is confirmed by a second measurement in a smaller RC in Section~\ref{sec:langevin}. Finally, a conclusion ends this paper.

\section{Theory}
\label{sec:theory}

The RC $Q$-factor is proportional to the ratio between the stored energy $U$ and the dissipated power $P_\mathrm{d}$ within the cavity as:

\begin{equation}
Q=\frac{\omega U}{P_\mathrm{d}}.
\label{Qdef}
\end{equation}

Let us consider an RC measurement setup where two arbitrary antennas are connected to a VNA in order to measure the RC $Q$-factor (from transmission coefficient). $N$ identical AUTs are located on masts within the working volume and unconnected to any measurement device. In this configuration, the RC $Q$-factor $Q_{\mathrm{L}x}$ can be decomposed as

\begin{equation}
Q^{-1}_{\mathrm{L}x}=Q^{-1}_\mathrm{c}+NQ^{-1}_{\mathrm{a,L}x}
\label{Q_L1}
\end{equation} 
where $Q_{\mathrm{a,L}x}$ is associated with the losses of a single AUT terminated by a load impedance $Z_{\mathrm{L}x}$ and $Q_\mathrm{c}$ is related to all other sources of losses (wall, transmitting and receiving antennas, stirrer, masts.). In order to isolate $Q_{\mathrm{a,L}x}$, one could perform an "empty" measurement where the $N$ AUTs are removed from the RC. Indeed, the difference between the $Q$-factors for the two measurements (with and without AUTs) leads to $NQ_{\mathrm{a,L}x}$. However, this requires two distinct measurements whose setup will eventually vary due to manual handling to remove the $N$ AUTs. This effect will be highlighted in section \ref{sec:langevin}. Thus, we suggest in this paper to perform two measurements, both with the $N$ AUTs located within the RC, but terminated by two different load impedances $Z_\mathrm{L1}$ and $Z_\mathrm{L2}$. The loads are modified thanks to remotely controlled switches and are identical for all AUTs. Hence, no manual handling is performed in this case. From  (\ref{Q_L1}), the difference between the $Q$-factors measured for two load impedances can be expressed as:
\begin{equation}
Q^{-1}_\mathrm{L1}-Q^{-1}_\mathrm{L2}=N\left(Q^{-1}_\mathrm{a,L1}-Q^{-1}_\mathrm{a,L2}\right)
\label{eq:Qdiff}
\end{equation}

The term $Q_\mathrm{c}$ vanishes in this difference as the setup remains identical. A first expression of the antenna $Q$-factor within an RC $Q_\mathrm{a}$, referred thereafter as eq1, has been derived by Hill et al. in \cite{Hill1998} such as:

\begin{equation}
\label{eq:q_a_hill}
Q_{\mathrm{a,L}x}^\mathrm{eq1} = \frac{Q^0_\mathrm{a}}{\eta_\mathrm{a}(1-|\Gamma_{\mathrm{a,L}x}|^{2})}
\end{equation} where $Q^0_\mathrm{a}=16\pi^2 V/\lambda^3$ is the $Q$-factor of an ideal lossless antenna, and $\Gamma_{\mathrm{a,L}x}$ is the reflection coefficient between the AUT input impedance and its load impedance $Z_{\mathrm{L}x}$. $\Gamma_{\mathrm{a,L}x}$ can be computed as \cite{Kurokawa1965}:

\begin{equation}
\label{eq:Gamma_L}
\left|\Gamma_{\mathrm{a,L}x}\right| = \left|\frac{\Gamma_{\mathrm{L}x}-\Gamma^{\dag}_\mathrm{a}}{1-\Gamma_{\mathrm{L}x}\Gamma_{\mathrm{a}}}\right|
\end{equation} where $\Gamma_{\mathrm{a}}$ and $\Gamma_{\mathrm{L}x}$ are the reflection coefficients of the AUT and the load impedance, respectively, according to $50~\Omega$. The superscript $^{\dag}$ stands for the complex conjugate. In 2018, Cozza \cite{Cozza2018} suggested modifying the $Q$-factor formula~(\ref{eq:q_a_hill}) in order to take into account the power reflected by the antenna load which is re-radiated towards the RC. The modified formula, referred then as eq2, is given as:

\begin{equation}
\label{eq:q_a_cozza}
Q_{\mathrm{a,L}x}^\mathrm{eq2} = \frac{Q^0_\mathrm{a}}{1-\eta^{2}_\mathrm{a}|\Gamma_{\mathrm{a,L}x}|^{2}}
\end{equation}  

Replacing $Q_\mathrm{a,L1}$ and $Q_\mathrm{a,L2}$ in (\ref{eq:Qdiff}) by their expressions from (\ref{eq:q_a_hill}) and (\ref{eq:q_a_cozza}), the AUT radiation efficiency $\eta_\mathrm{a}$ can be retrieved through two relations as:

\begin{equation}
\eta_\mathrm{a}^\mathrm{eq1}=\frac{Q^0_\mathrm{a}\left(Q^{-1}_\mathrm{L1}-Q^{-1}_\mathrm{L2}\right)}{N\left(\left|\Gamma_{\mathrm{a,L2}}\right|^2-\left|\Gamma_{\mathrm{a,L1}}\right|^2\right)}
\label{eq:radeff_hill}
\end{equation} 

and

\begin{equation}
\eta_\mathrm{a}^\mathrm{eq2}=\sqrt{\frac{Q^0_\mathrm{a}\left(Q^{-1}_\mathrm{L1}-Q^{-1}_\mathrm{L2}\right)}{N\left(\left|\Gamma_{\mathrm{a,L2}}\right|^2-\left|\Gamma_{\mathrm{a,L1}}\right|^2\right)}}.
\label{eq:radeff_cozza}
\end{equation}
It can be noticed that $\eta_\mathrm{a}^\mathrm{eq2}$ is the square root of $\eta_\mathrm{a}^\mathrm{eq1}$. The radiation efficiency estimation  $\eta_\mathrm{a}^\mathrm{eq2}$ using (\ref{eq:radeff_cozza})  is expected to be more accurate as it is based on the revised $Q$-factor formula from \cite{Cozza2018}, but the estimation  $\eta_\mathrm{a}^\mathrm{eq1}$ using (\ref{eq:radeff_hill})  will be evaluated in the following sections in order to highlight the limitations of the previous model from \cite{Hill1998}.

\section{Measurement}
\label{sec:esycom}
\subsection{Experiment}
\label{sec:experiment}

\begin{figure}[t!]
\centerline{\includegraphics[width=80mm,scale=1.5]{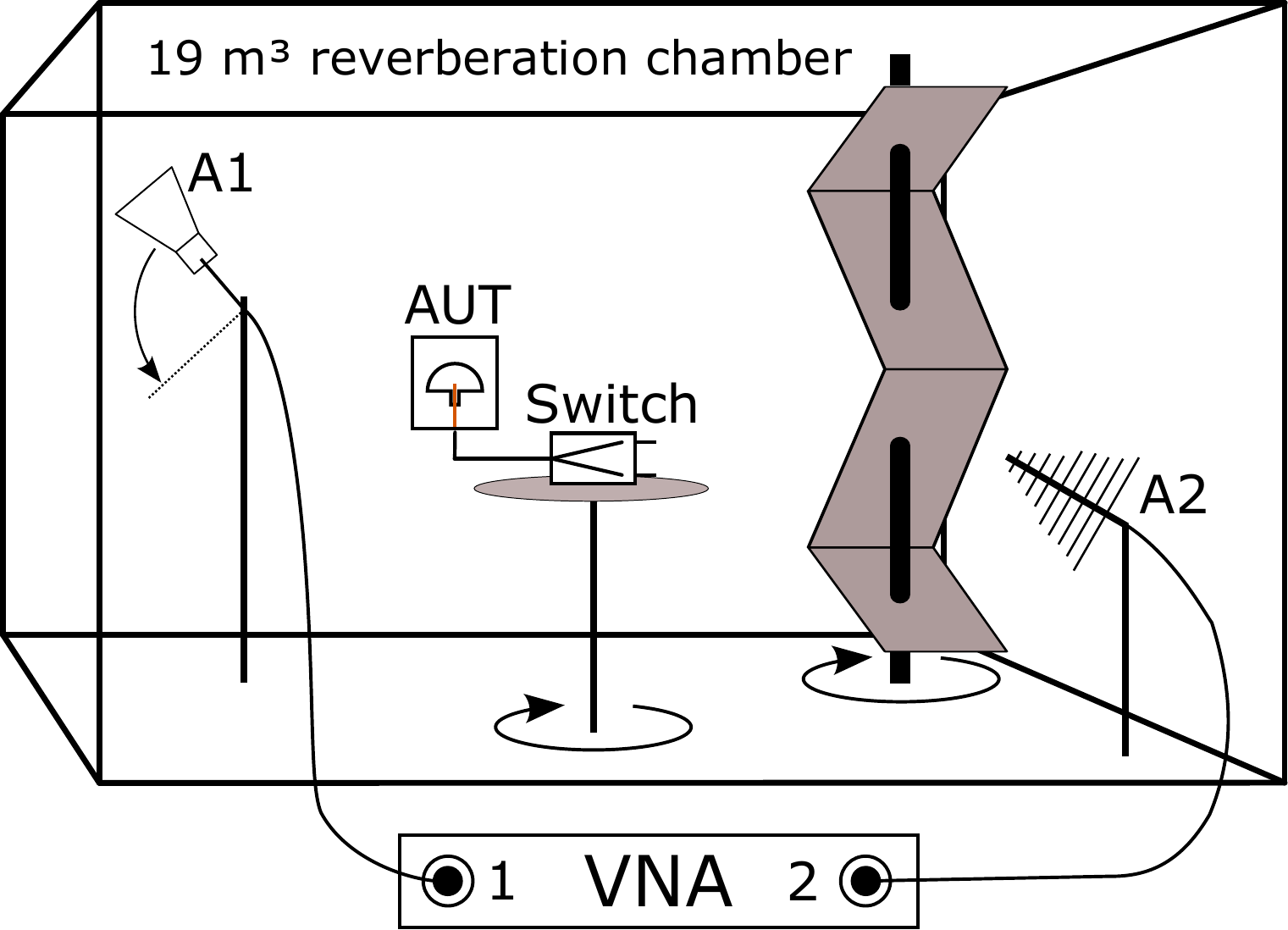}}
\caption{Antenna efficiency measurement setup in the ESYCOM RC. A1 and A2 are a horn antenna and a log-periodic antenna, respectively, and are connected to a VNA. A1 is oriented towards an edge of the RC and 17 different positions are considered. A2 is oriented towards the mode stirrer. The vertical mode stirrer rotates around its axis and 72 positions are considered over a revolution. The AUT is located on a mast and connected to a mechanical switch through a 30-cm cable. Four different orientations are considered for the AUT.}
\label{fig:setupEsycom}
\end{figure}

\begin{figure}[t!]
\centerline{\includegraphics[width=80mm,scale=1.5]{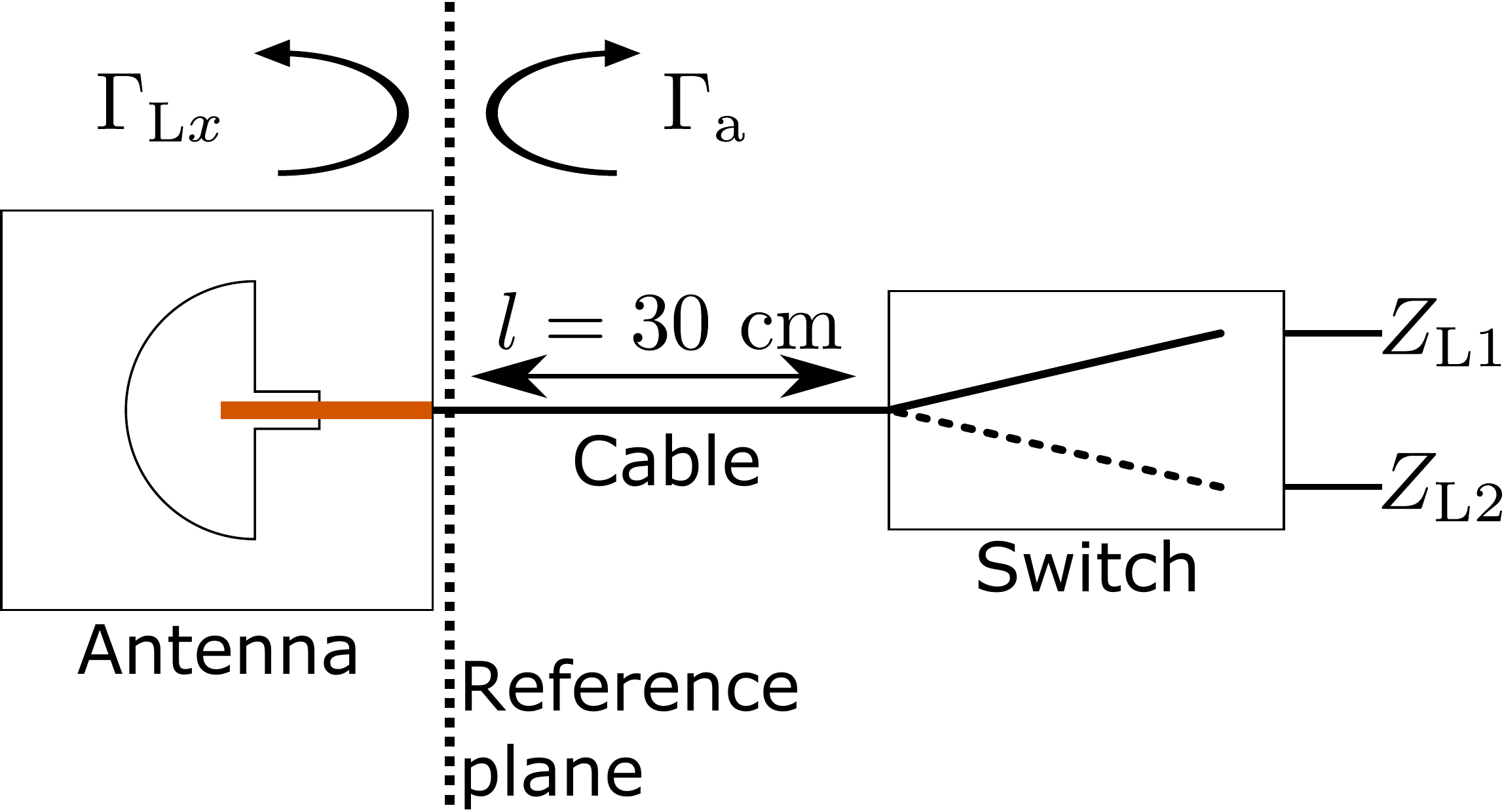}}
\caption{Reference plane considered for measuring both the antenna and the load reflection coefficients.}
\label{fig:calplane}
\end{figure}

The setup is presented in Fig.~\ref{fig:setupEsycom}. This experiment has been conducted at the ESYCOM laboratory in an RC whose dimensions are 2.95~m $\times$ 2.75~m $\times$ 2.35~m, equipped with a rotating metallic mode stirrer (72 equally spaced positions are considered). Two antennas (A1 and A2) are connected to a Rohde\&Schwarz ZNB20 vector network analyzer (VNA) in order to measure their scattering parameters in the  1.8~GHz to 2.8~GHz  frequency range (10001 frequency points). A1 is oriented towards an edge of the RC whereas A2 is oriented towards the mode stirrer. A1 is positioned on a vertically-rotating mast in order to perform source stirring over 17 positions. Both antennas are carefully positioned to avoid any direct coupling between them for all configurations;  indeed, A1 and A2 are orthogonally positioned and polarized.  The AUT is a wideband slot-based patch antenna (details in \cite{Krouka2020}) that is connected, through a 30-cm long coaxial cable, to a remotely-controlled mechanical switch that allows switching between two load impedances (Fig.~\ref{fig:calplane}). The AUT and the switch are positioned on a rotating mast and four positions are considered to perform the AUT stirring. In order to maximize the difference between the two measured $Q$-factors, and thus, the measurement sensitivity, an open circuit (OC) and a 50~$\Omega$ load are chosen as the two load impedances $Z_\mathrm{L1}$ and $Z_\mathrm{L2}$, respectively. 
All measurements are performed in a single run, without manual manipulation.

\begin{figure}[t!]
\centerline{\includegraphics[width=90mm,scale=1.5]{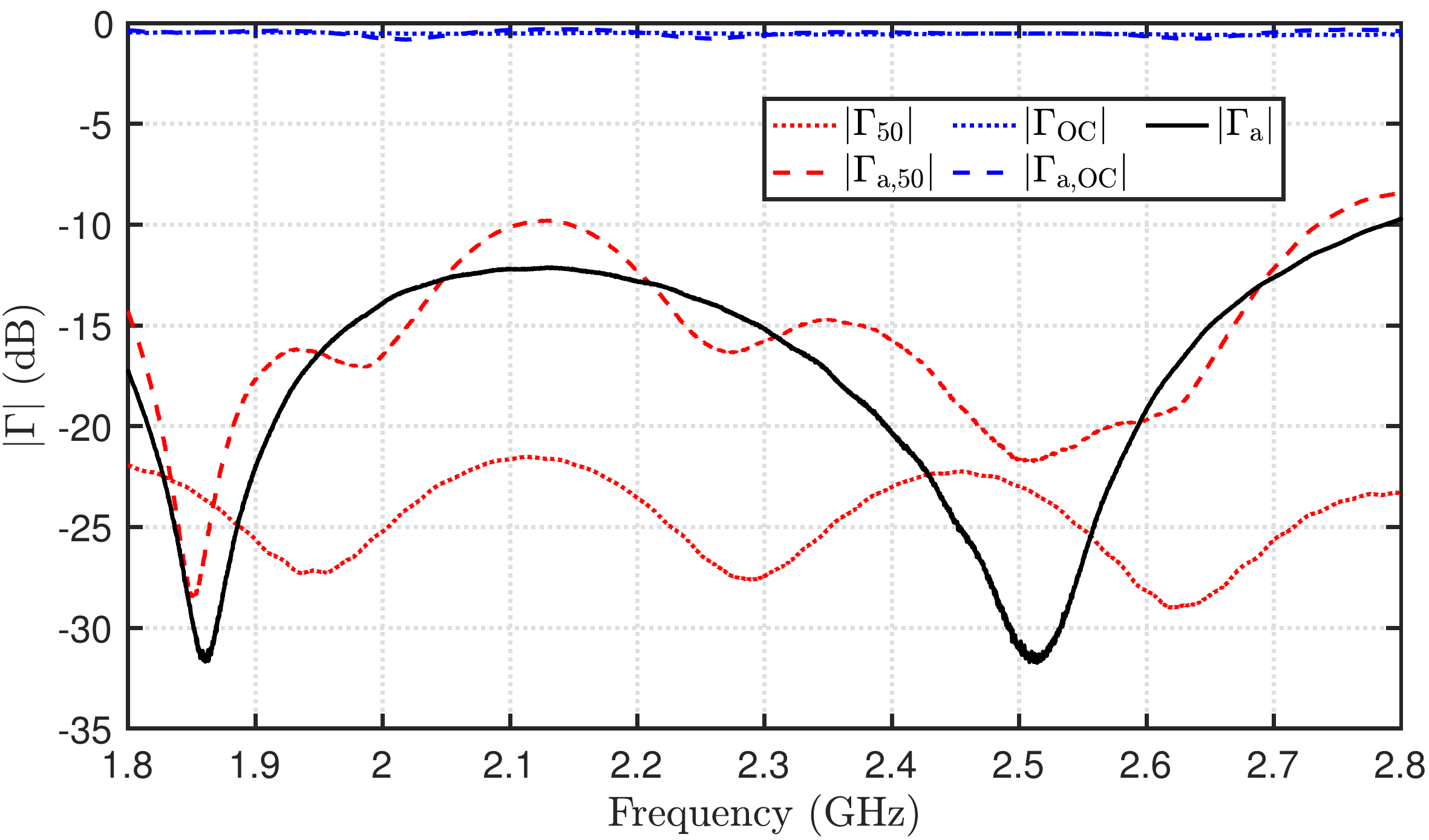}}
\caption{AUT and load impedances reflection coefficients as a function of frequency.}
\label{fig:Gamma}
\end{figure}

To assess the validity of (\ref{eq:radeff_hill}) and (\ref{eq:radeff_cozza}), we first perform an invasive estimation of $\left|\Gamma_{\mathrm{a,L}x}\right|$. With a VNA, we measure on the one hand the AUT reflection coefficient $\Gamma_{\mathrm{a}}$ within an AC, and on the other hand the reflection coefficients $\Gamma_\mathrm{L1}=\Gamma_\mathrm{OC}$ and $\Gamma_\mathrm{L2}=\Gamma_\mathrm{50}$, associated to the two load impedances $Z_\mathrm{L1}$ and $Z_\mathrm{L2}$. As shown in Fig.~\ref{fig:calplane}, the measurement reference plane is located at the antenna's port so that the losses associated to the cable and the switch are included in $\Gamma_\mathrm{OC}$ and $\Gamma_\mathrm{50}$. From this 2-step measurement, $\left|\Gamma_{\mathrm{a,OC}}\right|$ and $\left|\Gamma_{\mathrm{a,50}}\right|$ are deduced from (\ref{eq:Gamma_L}) and results are presented in Fig.~\ref{fig:Gamma} as a function of the frequency.
We can notice that the AUT is well matched to 50~$\Omega$ in the whole frequency range ($\left|\Gamma_{\mathrm{a}}\right|$ less than -10~dB). Due to the switch insertion losses as well as the losses due to the cable between the AUT and the switch,  $\left|\Gamma_{\mathrm{OC}}\right|$ is not equal to 0~dB but its average over the frequency band is of $-0.51$~dB. $\left|\Gamma_{\mathrm{50}}\right|$ is less than -20~dB; thus, this load acts as a good matched load even though the presence of the switch and the cable. As the AUT is well matched, $\left|\Gamma_{\mathrm{a,50}}\right|$ and $\left|\Gamma_{\mathrm{a,OC}}\right|$ exhibit similar behaviors to $\left|\Gamma_{\mathrm{a}}\right|$ and $\left|\Gamma_{\mathrm{OC}}\right|$, respectively.

\begin{figure}[t!]
\centerline{\includegraphics[width=90mm,scale=1.5]{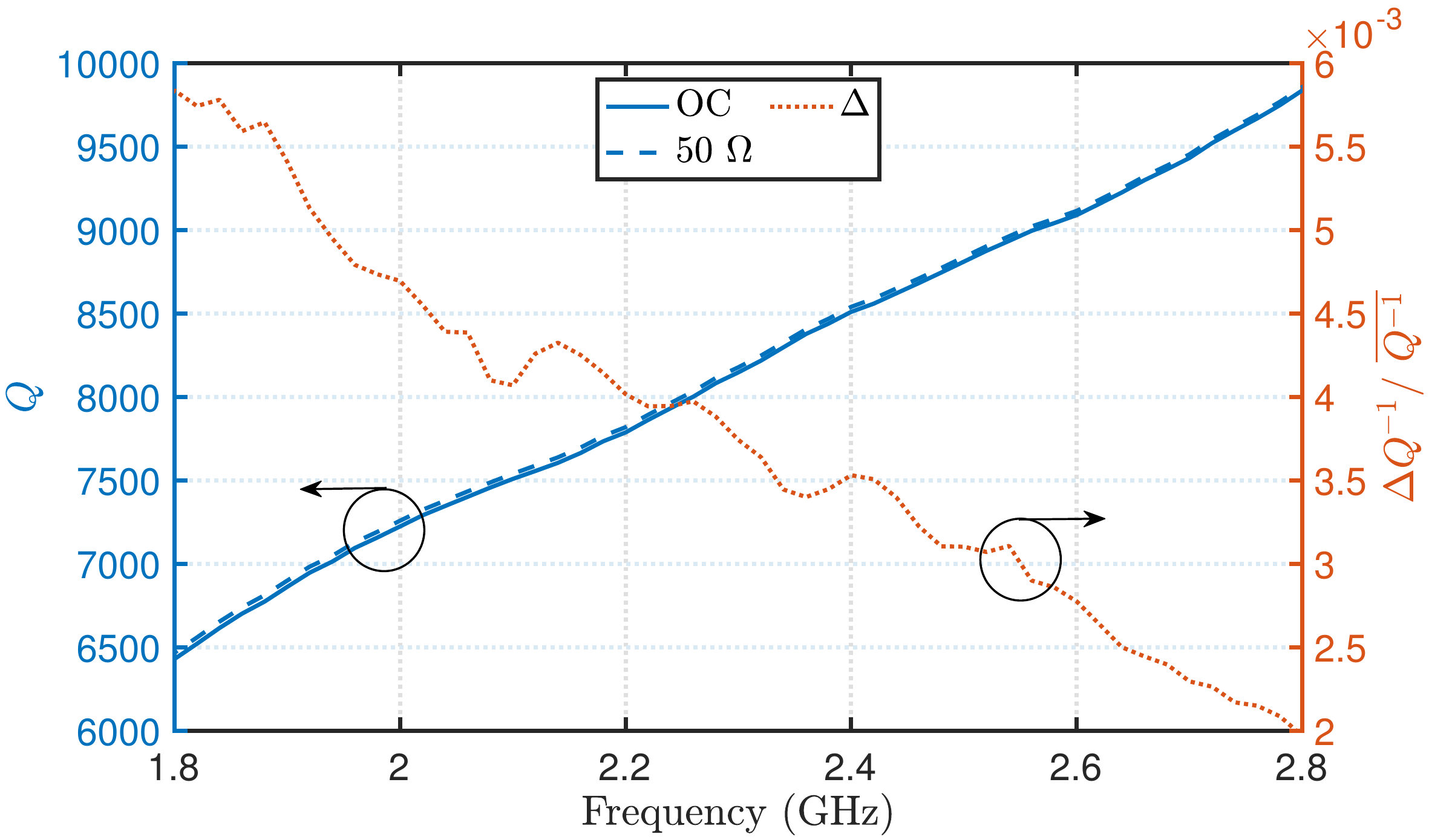}}
\caption{$Q$ as a function of frequency (left) and $\Delta Q^{-1}/\overline{Q^{-1}}$ as a function of frequency (right).}
\label{fig:Qesycom}
\end{figure}

RC $Q$-factors for both load conditions are computed in the time domain using the transmission coefficient $S_\mathrm{21}$ between both measurement antennas (A1 and A2) by fitting the power delay profile $\mathrm{PDP}=\left<\left|\mathrm{IFT}(S_\mathrm{21})\right|^2\right>$ \cite{Holloway2012} where the average denoted $\left<.\right>$ is calculated over all the configurations in addition to a sliding frequency window of 200~MHz. As the two $Q$-factors are very close, we present also $\Delta Q^{-1}/\overline{Q^{-1}}=2\left(Q^{-1}_{50}-Q^{-1}_\mathrm{OC}\right)/\left(Q^{-1}_{50}+Q^{-1}_\mathrm{OC}\right)$ in Fig.~\ref{fig:Qesycom} as a function of frequency. This quantity is always positive, meaning that $Q^{-1}_\mathrm{OC}<Q^{-1}_{50}$ for all frequencies. This is coherent as more losses are expected within the RC when the AUT is connected to the 50~$\Omega$ load. Also, it decreases as a function of frequency, which is expected as the contribution of the AUT on the RC $Q$-factor becomes smaller.


\subsection{Results}
\label{sec:results}

\begin{figure}[t!]
\centerline{\includegraphics[width=90mm,scale=1.5]{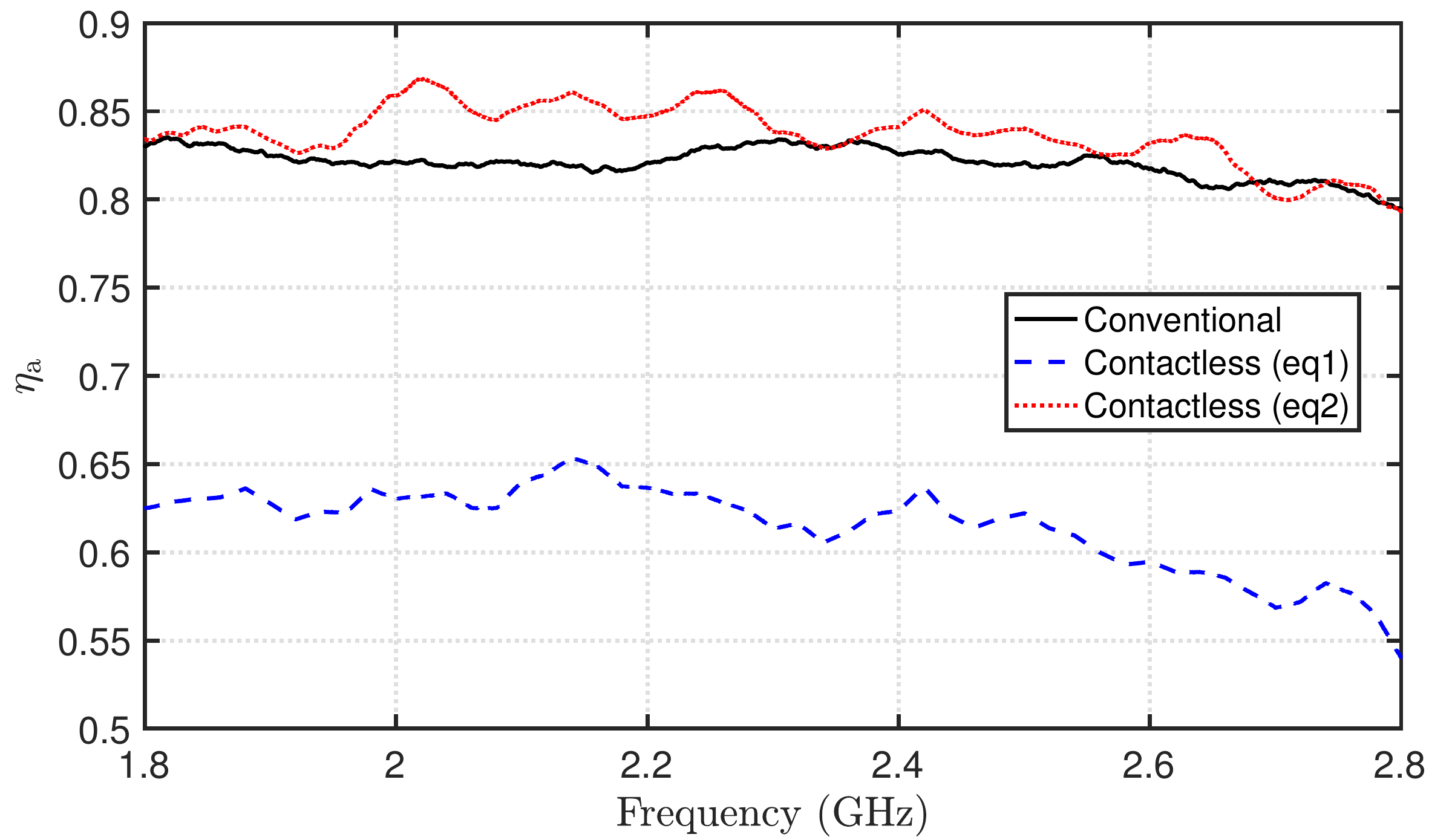}}
\caption{Radiation efficiency of the patch antenna measured in the ESYCOM RC using the invasive and the contactless approach, as a function of frequency. Results are smoothed over a 30~MHz window.}
\label{fig:eff_sans_contact_hill}
\end{figure}

The radiation efficiency measured using the contactless setup is presented in Fig.~\ref{fig:eff_sans_contact_hill} using eq1 (\ref{eq:radeff_hill}) and eq2 (\ref{eq:radeff_cozza}) as a function of frequency. To evaluate the radiation efficiency accuracy, results are compared to those obtained through a conventional invasive measurement method within the same RC (\textit{two-antenna approach} in \cite{Holloway2012}). 

This method is chosen as a reference as it is widely used and well known to provide accurate and repeatable radiation efficiency estimations within an RC \cite{Hubrechsen2021, Bronckers2020}.

These results show that the newly introduced method using (\ref{eq:radeff_cozza}), i.e., eq2, allows obtaining very similar results to the ones obtained using a conventional invasive RC approach applied in the same measurement conditions (same RC $Q$-factor and same stirring processes). Indeed, the averaged relative error is equal to only 2.3~\%. 

As the estimated efficiency using the conventional method is almost constant over the frequency range, we compute the mean and the standard deviation of the three estimated efficiencies for comparison purposes; they are presented in Table~\ref{tab:stat}.  The radiation efficiency averaged over the whole frequency band is equal to 81.9~\% with the conventional method and 83.7~\% with the contactless approach in the case of eq2. However, the standard deviation is higher for the contactless approach. 

Indeed, the proposed method is based on the estimation of a small $Q$-factor variation due to the AUT load modification and is therefore more sensitive to the non-ideal behavior of the RC, leading to higher fluctuations as a function of frequency.

It has to be noted that, as expected, the contactless approach using (\ref{eq:radeff_hill}), i.e., eq1, leads to a strongly underestimated radiation efficiency (61.6~\% on average over the frequency bandwidth). This is due to the fact that (\ref{eq:radeff_hill}) does not take into account that a part of the power arising on the AUT is actually reflecting back into the RC (due to mismatches). This power is thus considered as absorbed by the AUT, which artificially decreases its radiation efficiency. These results demonstrate the possibility to estimate the antenna radiation efficiency using a contactless approach, i.e., without connecting the AUT to the VNA, by measuring the RC $Q$-factor for two AUT load conditions. Also, it confirms that the antenna $Q$-factor formula introduced in \cite{Cozza2018} better models the antenna absorption within an RC.

\begin{table}[t!]
\centering
\caption{Radiation Efficiency Mean Value and  Standard Deviation  over the Frequency Range for the Experiments Conducted in the 19~m$^3$ Reverberation Chamber.}
\setlength{\tabcolsep}{5pt}
\begin{tabular}{| l | c | c |}

\hline
 & Mean  &  Standard Deviation  \\
\hline
Conventional & $81.9$~\% &  $0.0102$\\
\hline
Contactless eq1& $61.6$~\%  &  $0.0228$\\
\hline
Contactless eq2 & $83.7$~\% &  $0.0163$\\
\hline
Contactless eq2 (ideally matched AUT)& $82.3$~\% &  $0.0149$\\
\hline
Contactless eq2 (ideal loads) &  $78.6$~\% & $0.0131$\\
\hline
\end{tabular}
\label{tab:stat}
\end{table}

\subsection{Reflection coefficient approximations}
\label{sec:approx}

\begin{figure}[t!]
\centerline{\includegraphics[width=90mm,scale=1.5]{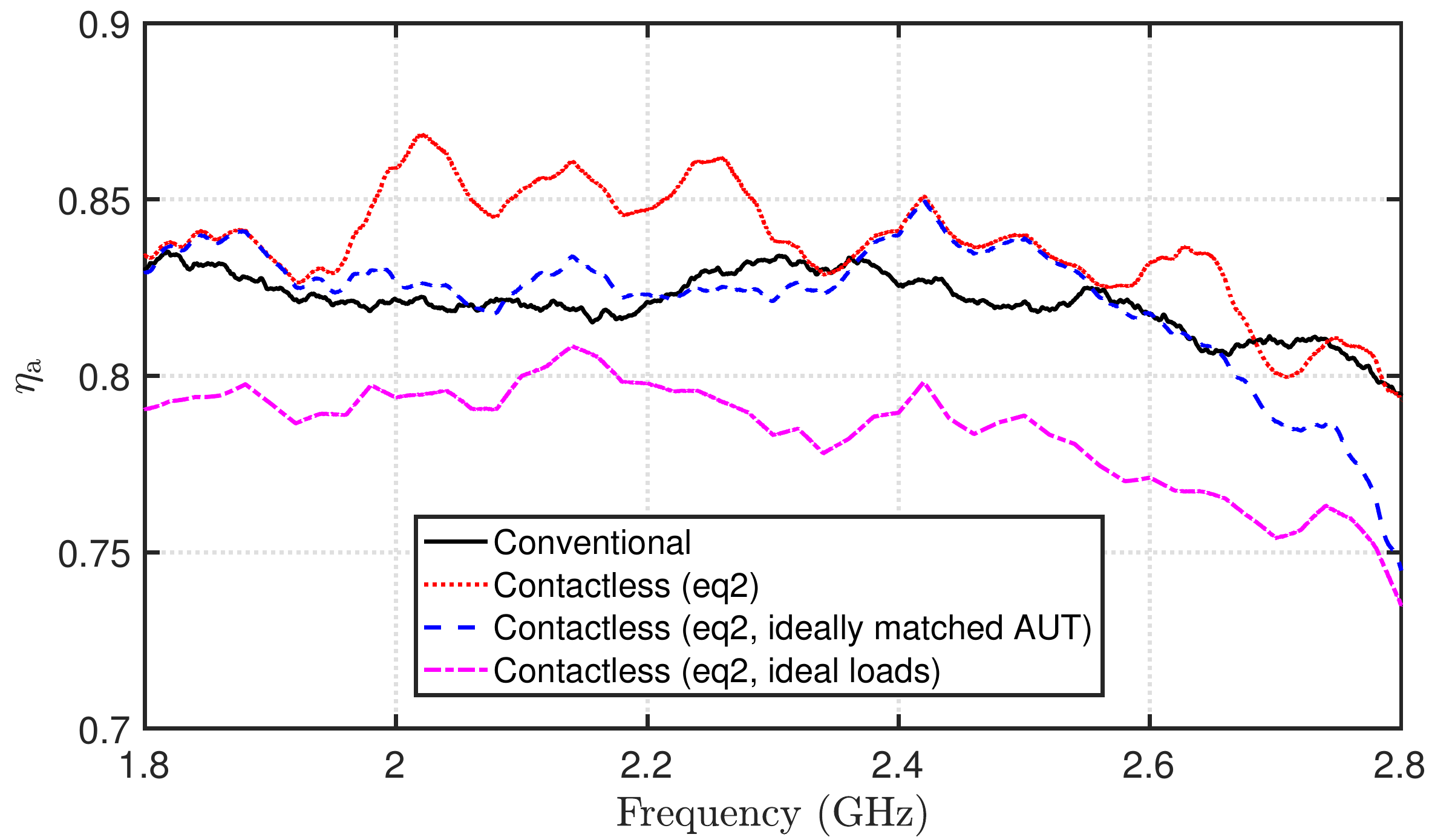}}
\caption{AUT radiation efficiency as a function of frequency, measured using the contactless approach considering different approximations regarding the reflection coefficients.}
\label{fig:eff_approx}
\end{figure}

The contactless method requires the determination of $\left|\Gamma_{\mathrm{a,L}x}\right|$ from (\ref{eq:Gamma_L}) and thus the measurement of both reflection coefficients $\Gamma_{\mathrm{a}}$ and $\Gamma_{\mathrm{L}x}$. In this paper, they are measured directly using a VNA, i.e., using an invasive measurement. Although the impedance is usually less sensitive than radiation properties to the proximity of cables, impedance measurements might not be possible from a practical point of view, especially if the AUT is specifically designed including an integrated miniaturized switching device. Thus, we evaluate in the following the impact of no \textit{a priori} knowledge on these reflection coefficients. Accordingly, two approximations are successively done: (1) the AUT is assumed to be ideally matched to 50~$\Omega$ so that $\Gamma_{\mathrm{a}}=0$ and thus $\left|\Gamma_{\mathrm{a,L}x}\right| = \left|\Gamma_{\mathrm{L}x}\right|$, and (2) we consider an ideal OC so that $\Gamma_{\mathrm{OC}}=1$ and thus $\left|\Gamma_{\mathrm{a,OC}}\right|= 1$ and an ideal 50~$\Omega$ load so that $\Gamma_{50} = 0$ and thus $\left|\Gamma_{\mathrm{a,50}}\right|=\left|\Gamma_{\mathrm{a}}\right|$. Radiation efficiencies obtained from these two approximations are presented in Fig.~\ref{fig:eff_approx} and compared with the conventional method and the contactless one without approximation. Considering the AUT ideally matched to 50~$\Omega$ leads to very similar results to the ones without approximation (the average radiation efficiency is equal to 82.3~\%) which is due to the fact that the AUT is well-matched in this frequency band (see Fig.~\ref{fig:Gamma}). Indeed, when $\Gamma_{\mathrm{a}}$ is very low (around 1.85~GHz and 2.5~GHz), the two curves merge. However, if we consider ideal loads, a systematic underestimation of the radiation efficiency is observed (the average radiation efficiency is equal to 78.6~\%). This is mainly due to the switch and the cable insertion losses located between the calibration port and the load, which are not taken into account and thus, not properly compensated.

\subsection{Stirring effect}
\label{sec:stirring}
This part is dedicated to the evaluation of the measurement uncertainties according to the stirring process. All results shown in the previous parts were obtained by applying all the stirring possibilities of the measurement setup, including the mechanical stirring (72 configurations), the source stirring (17 configurations) and the AUT stirring (4 configurations), leading up to 4896 configurations. The objective of such stirring process is to lower the measurement uncertainties, i.e., to lower the standard deviation of the radiation efficiency estimation. However, the accuracy of any RC measurements relies on the effective sample size generated by the stirring process, i.e., the number of uncorrelated configurations. 
A correlation coefficient $\rho$ is first calculated using the 72 stirrer positions for a single source position and a single AUT position, based on the transmission coefficient between the two measurement antennas. The effective sample size $N_\mathrm{eff}$ is then deduced from $\rho$ \cite{Lemoine2008}.
Since it presents fluctuations as a function of frequency, its value is averaged in the  2~GHz to 2.4~GHz  frequency band. This estimation is made for each source and AUT position before being averaged and leads to $N_\mathrm{eff}=59$. It has to be noted that $N_\mathrm{eff}$ has been computed for the two AUT load conditions but it leads to very similar results thus, only the results obtained for the OC case are presented thereafter.

Once the effective sample size is known, it is possible to evaluate the normalized standard deviation of the radiation efficiency estimation as \cite{Chen2013}:
\begin{equation}
\frac{\sigma_{\eta_\mathrm{a}}}{\left<\eta_\mathrm{a}\right>}\approx\sqrt{\frac{2}{N_\mathrm{eff}}}.
\label{eq:std}
\end{equation}
In the present case ($N_\mathrm{eff}=59$), the normalized standard deviation is estimated to be equal to 0.184. In order to validate this result, we now aim at evaluating the normalized standard deviation of the radiation efficiency estimation from the measurement results. As two types of stirring, namely the source stirring and the AUT stirring, are performed in addition to the conventional mechanical stirring, two different estimations of the normalized standard deviation are computed: one using the source stirring and the other one using the AUT stirring. First, we estimate 
$\sigma_{\eta_\mathrm{a}}^{p_\mathrm{src}}$ from all source positions and for one AUT position. The result is then averaged over all AUT positions $\left<\sigma_{\eta_\mathrm{a}}^{p_\mathrm{src}}\right>_{p_\mathrm{AUT}}$. The same process is made considering only the 4 AUT positions for one source position, before averaging over all source positions $\left<\sigma_{\eta_\mathrm{a}}^{p_\mathrm{AUT}}\right>_{p_\mathrm{src}}$. Results are presented in Table~\ref{tab:sigma}. Both estimations from the measurement results are in very good agreement with the theoretical prediction (relative error of 6~\% for the source stirring estimation and 1~\% for the AUT stirring).

This confirms that the normalized standard deviation  can be predicted directly from $N_\mathrm{eff}$. In the present case, we evaluated 10 uncorrelated source positions out of the 17 using the same process than the one used for the mechanical stirring. If we consider that the 4 AUT positions are uncorrelated (4 positions being too low to accurately estimate the effective sample size), it comes $4 \times 10 \times 59 =2360$ uncorrelated configurations. According to (\ref{eq:std}), this leads to a predicted normalized standard deviation that is equal to 0.029.

\begin{table}[t!]
\centering
\caption{Normalized Standard Deviations of the Estimated Radiation Efficiency According to the Mechanical Stirring Process. }
\setlength{\tabcolsep}{5pt}
\begin{tabular}{| l | c |}
\hline
 &  Normalized Standard Deviation \\
\hline
Prediction \cite{Chen2013} & 0.184 \\
\hline
Source stirring & 0.173\\
\hline
AUT stirring & 0.186\\
\hline
\end{tabular}
\label{tab:sigma}
\end{table}


\section{Validation in a Small RC}
\label{sec:langevin}
\begin{figure}[t!]
\centerline{\includegraphics[width=80mm,scale=1.5]{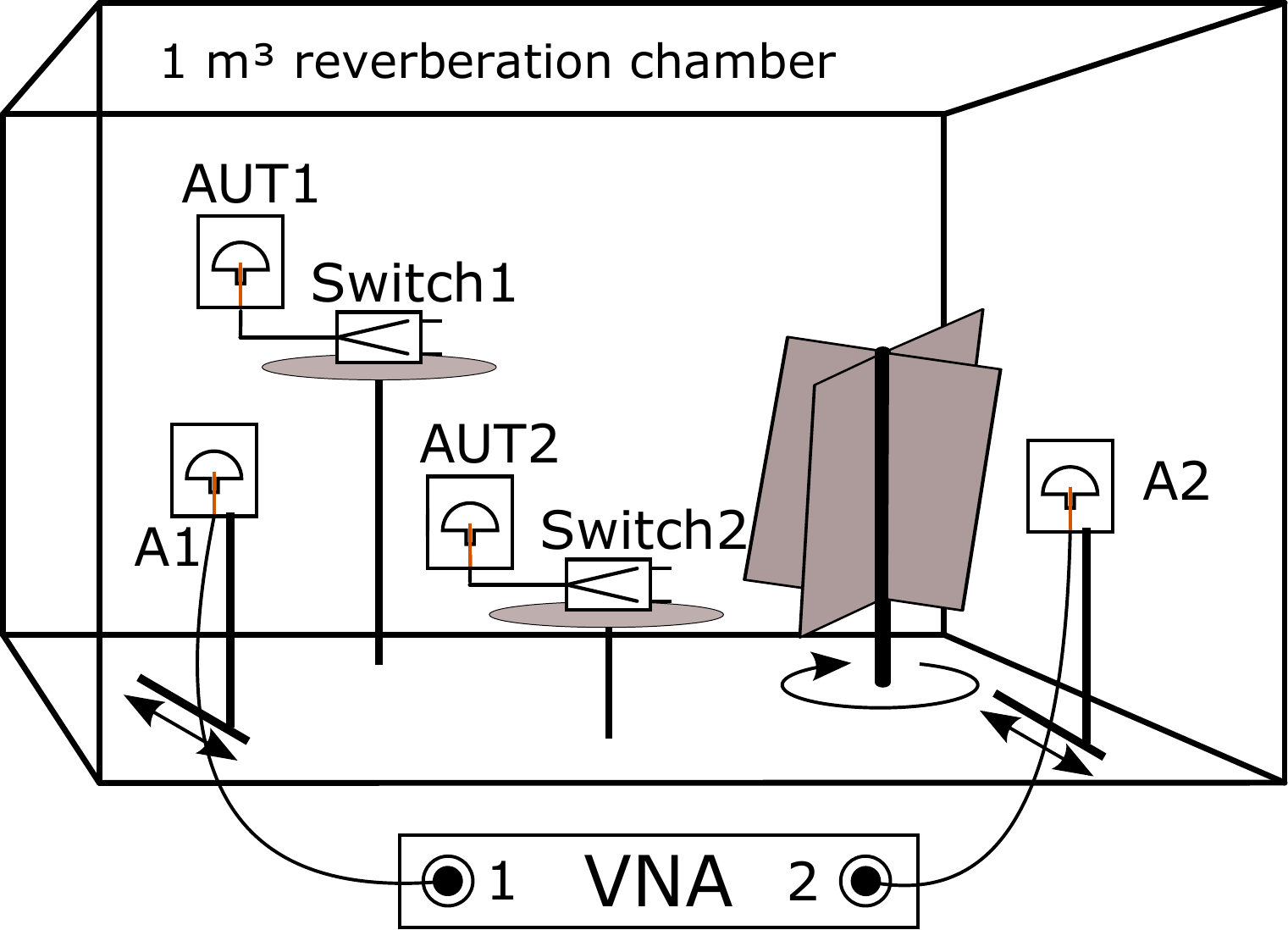}}
\caption{Contactless antenna efficiency measurement setup in the small RC. The vertical mode stirrer is rotating around its axis and both measurement antennas A1 and A2 are translated along linear stages.}
\label{fig:setup}
\end{figure}

In this section, we aim at validating the proposed measurement method using a much smaller RC, thus exhibiting a much smaller $Q$-factor. These experiments have been conducted in the 1~m$^3$ RC of the Institut Langevin with the experimental setup presented in Fig.~\ref{fig:setup}. The AUT is the same wideband patch antenna as in the previous section. The $Q$-factor being much smaller than in the large RC (about 1700 at the central frequency), two identical AUTs have been placed within the RC in order to enhance the AUT absorption strength so $N=2$ in (\ref{eq:radeff_cozza}). They are both connected to their own mechanical switch (that can switch between an OC and a 50~$\Omega$ load) through a 30-cm coaxial cable. For practical reasons (especially the limited size of the RC), the two antennas A1 and A2 are the same as the AUTs so that four identical patch antennas are used: two as AUTs and two as measurement antennas (A1 and A2). Scattering parameters are measured thanks to an Anritsu VNA in the same frequency range  (1.8~GHz to 2.8~GHz)  with 10001 frequency points. A1 and A2 antennas are positioned on a translation stage in order to perform source stirring over 9 configurations (3 positions for each antenna) in addition to the mechanical rotating stirrer that provides 72 positions. Thus, all results are averaged over 648 configurations. 

\begin{figure}[t!]
\centerline{\includegraphics[width=90mm,scale=1.5]{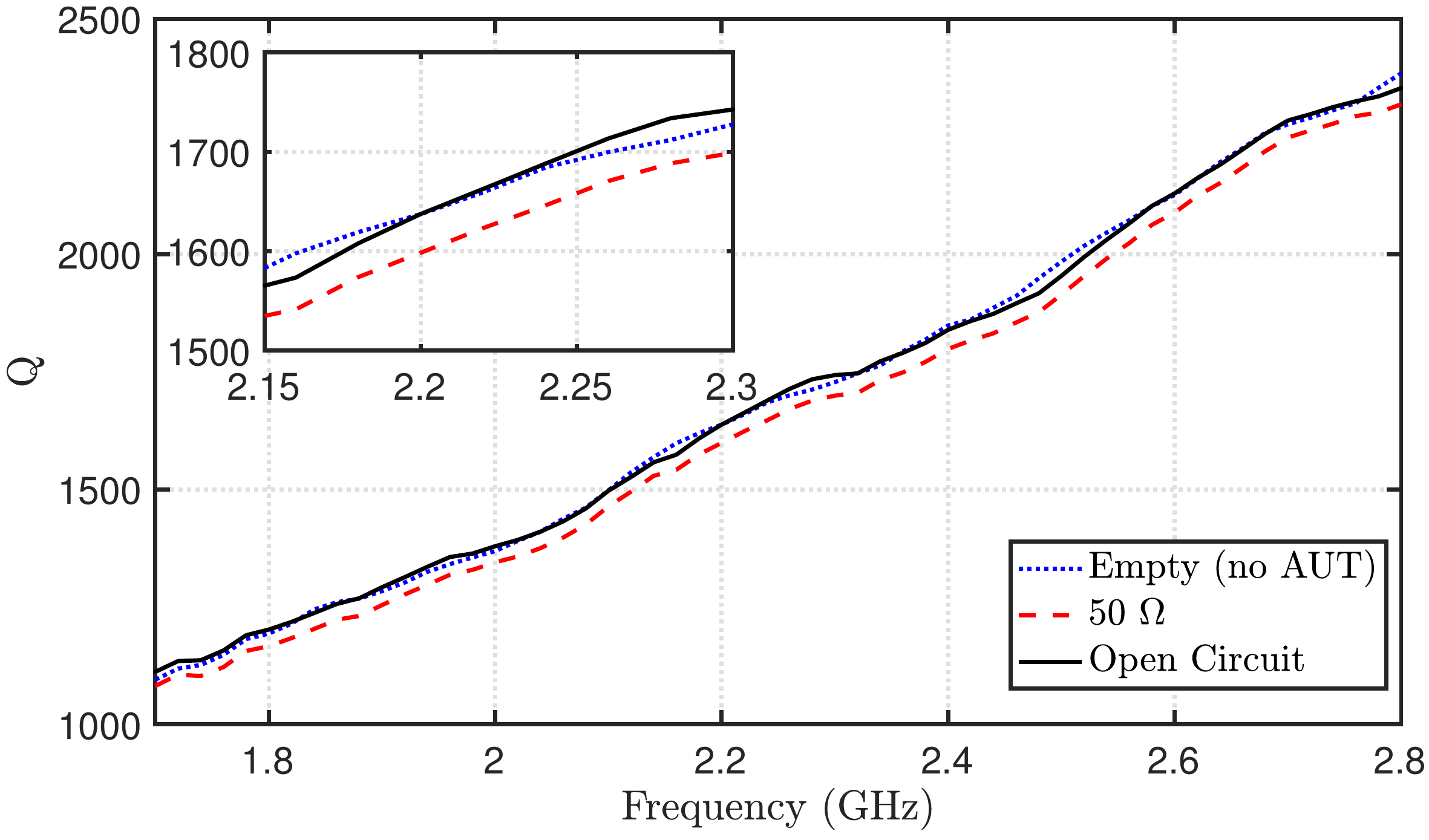}}
\caption{RC $Q$-factors of the small RC as a function of frequency, measured without any AUT (blue) or when both AUTs are connected to an OC (black) and a 50~$\Omega$ load (red).}
\label{fig:Q}
\end{figure}

The same procedure as in the previous section is conducted here regarding the AUT and the load impedances reflection coefficients measurement, the results are not presented for brevity. $Q$-factors are presented in Fig.~\ref{fig:Q} as a function of frequency. As expected, $Q$ increases with frequency, and it is slighty lower for the 50~$\Omega$ case as more energy is absorbed by this load. In order to highlight the sensitivity of the $Q$-factor estimation, we also computed the $Q$-factor for the empty case, i.e. when the AUTs are removed, whereas everything else, including the cables and the switches, are kept within the RC. We can see that this result is not consistent with the previous ones. Indeed, due to the estimation fluctuations, $Q$ in the empty case is sometimes higher and sometimes lower than the one with AUT connected to the OC. This is precisely why we suggested in this approach the use of two load conditions obtained without manual manipulation and no empty RC measurement.

\begin{figure}[t!]
\centerline{\includegraphics[width=90mm,scale=1.5]{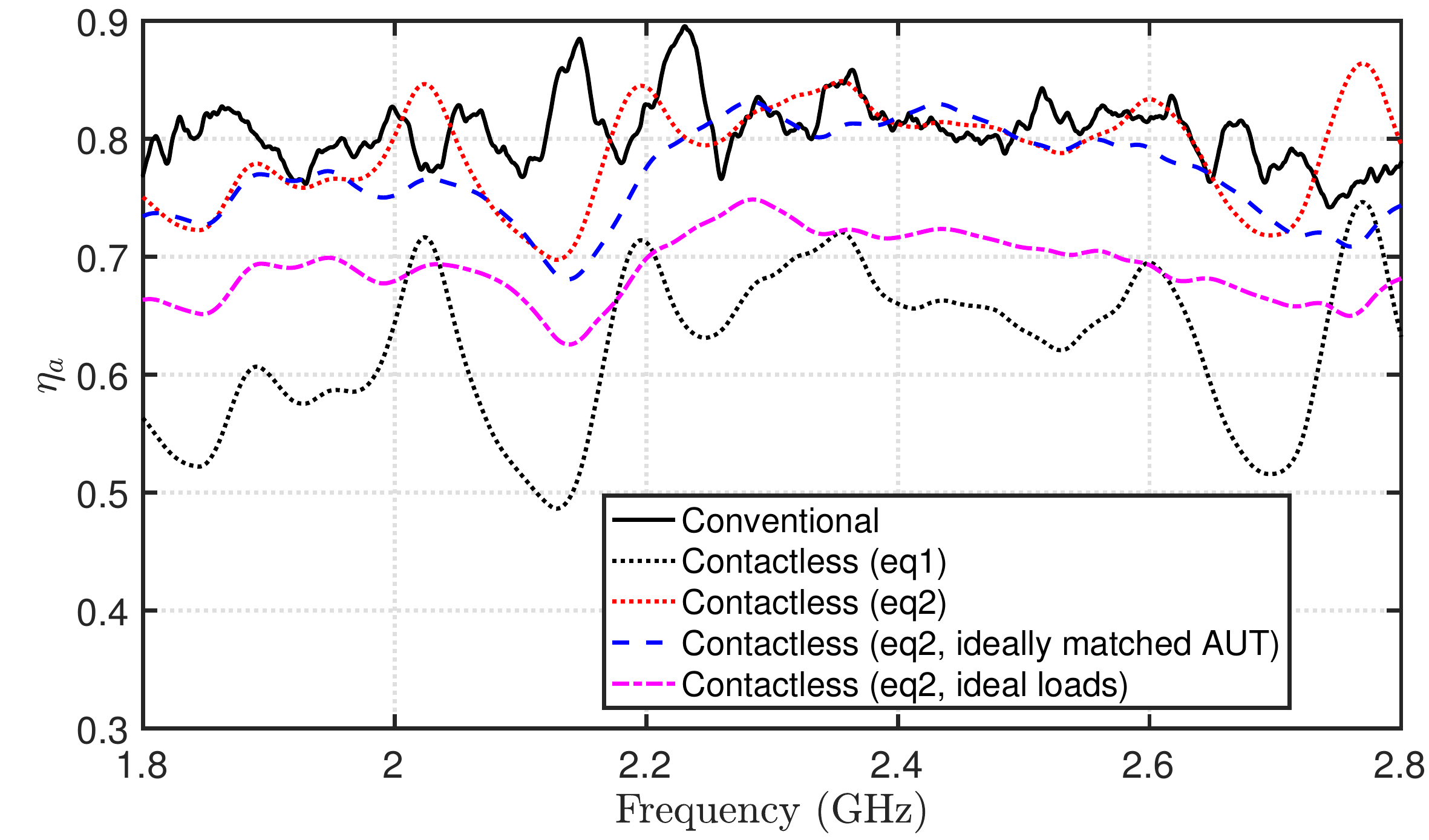}}
\caption{AUT radiation efficiency as a function of frequency, measured in the small RC using the invasive and the contactless approaches,  considering different approximations regarding
the reflection coefficients. The results are smoothed over a 30~MHz window.}
\label{fig:eff_langevin_approx}
\end{figure}

The radiation efficiency is measured using the conventional setup as well as with the contactless setup and results are presented in Fig.~\ref{fig:eff_langevin_approx} as a function of frequency. The values of $\left|\Gamma_{\mathrm{a,L}x}\right|$ have been computed by averaging the measurements performed on the 2 switches. Mean values as well as standard deviations  over the frequency range  for each estimation are presented in Table~\ref{tab:stat2}. These results confirm those obtained in section~\ref{sec:results}: a very similar radiation efficiency estimation is obtained between the conventional and the contactless (eq2) methods (2~\% absolute difference on average, and a relative difference of 6~\% on average) although the standar deviation as a function of frequency is higher for the contactless case. It has to be noted that the mean number of uncorrelated stirrer positions, between  2~GHz and 2.4~GHz,  is evaluated to be 21, for a mean number of uncorrelated measurement antenna configurations of 5.4. This leads to 113 uncorrelated samples, which is much smaller than the 2360 uncorrelated samples of the ESYCOM RC; this explains why the standard deviations are larger than the ones obtained previously. Again, the radiation efficiency estimated through eq1 is not consistent with the one obtained using the conventional method. Also, the same approximations as the ones made in \ref{sec:approx} regarding the reflection coefficients are made. Considering the AUT perfectly matched to 50~$\Omega$ leads to a slight underestimation of only 2~\% on average whereas considering ideal loads leads to a strong underestimation of about 10~\%.


\begin{table}[t!]
\centering
\caption{Radiation Efficiency Mean Value and  Standard Deviation  over the Frequency Range for the Experiments Conducted in the 1~m$^3$ Reverberation Chamber.}
\setlength{\tabcolsep}{5pt}
\begin{tabular}{| l | c | c |}
\hline
 & Mean &  Standard Deviation  \\
\hline
Conventional & $80.9$~\% &   $0.0254$ \\
\hline
Contactless eq2 & $78.9$~\% &   $0.0424$ \\
\hline
Contactless eq2 (ideally matched AUT)& $76.9$~\% &  $0.0387$ \\
\hline
Contactless eq2 (ideal loads) & $69.0$~\% &  $0.0278$ \\
\hline
\end{tabular}
\label{tab:stat2}
\end{table}

\section{Conclusion and Discussion}

In this paper, we introduced an original measurement method to evaluate the antenna radiation efficiency within an RC.  It is contactless so that the method does not require to connect the AUT to an analyzer.  It is based on the measurement of the RC $Q$-factor for two load conditions of the AUT. It is emphasized that the two measurements are performed without modifying the measurement setup (no manual handling) thanks to remotely-controlled mechanical switches. An OC and a 50~$\Omega$ load have been considered here as the two load impedances in order to enhance the measurement sensitivity. It has been validated with a wideband patch antenna in the  1.8~GHz to 2.8~GHz  frequency range in two different RCs of different volumes: 1~m$^3$ and 19~m$^3$. The retrieved radiation efficiency is very similar to the one obtained using a conventional invasive approach \cite{Holloway2012} (average relative difference of 2.3~\% in the large RC and 6~\% in the small one). 


This method paves the way for non-invasive antenna radiation efficiency measurement, which is highly suitable for electrically small antennas as an invasive setup disturbs their impedance and radiation properties. However, this method also exhibits limitations. The contactless approach is based on the estimation of the RC $Q$-factor variation for two AUT load conditions. Depending of the considered setup (AUT, loads, and RC $Q$-factor), this variation may be small. Therefore, the contactless method is more sensitive to the non-ideal behavior of the RC and its stirring capabilities than conventional measurements. This has been seen in the second set of measurement (Section \ref{sec:langevin}) in an RC with a smaller $Q$-factor: the retrieved efficiency exhibits higher fluctuations over the frequency range than in the RC of higher $Q$-factor. This limitation can be counterbalance by performing additional stirring as well as enhancing the number of AUTs (increasing $N$) in order to raise the RC $Q$-factor variation.


In this paper, we also confirmed that the antenna $Q$-factor formulation introduced in 2018 \cite{Cozza2018} describes the absorption introduced by an antenna within an RC in a better manner than the former model introduced by D. Hill in 1998 \cite{Hill1998}. Indeed, some of the energy reflected back by the antenna load towards the RC was not taken into account by the original model, leading to a strong radiation efficiency underestimation once using this model in our method.

Future works will include miniaturization of the switching device in order to integrate it into the AUT. Specific antenna design including the switching device could be realized for testing purposes only.


%

\appendices




\ifCLASSOPTIONcaptionsoff
  \newpage
\fi

\end{document}